\documentclass[twocolumn]{aastex63}
\usepackage{amsmath}

\usepackage{graphicx}
\usepackage{amsmath}
\usepackage{mathtools}
\usepackage{kantlipsum}
\usepackage{xparse}
\usepackage{enumerate}
\usepackage[version=3]{mhchem}
\usepackage{amsmath,array}
\usepackage[T1]{fontenc}
\usepackage{courier}
\usepackage{booktabs}

\usepackage{sidecap}
\usepackage{tabularx}

\usepackage{fancyhdr}
\submitjournal{AJ}

\graphicspath{{./}{figures/}}

\begin{document}

\title{A Non-Detection of Iron in the First High-Resolution Emission Study of the Lava Planet 55 Cnc e}

\author[0000-0002-0470-0800]{Kaitlin C. Rasmussen}
\altaffiliation{co-first author}
\affiliation{Department of Astronomy and Astrobiology Program, University of Washington, Box 351580, Seattle, Washington 98195, USA}
\affiliation{NASA Nexus for Exoplanet System Science, Virtual Planetary Laboratory Team, Box 351580, University of Washington, Seattle, Washington 98195, USA}

\author[0000-0003-3429-4142]{Miles H. Currie}
\altaffiliation{co-first author}
\affiliation{Department of Astronomy and Astrobiology Program, University of Washington, Box 351580, Seattle, Washington 98195, USA}
\affiliation{NASA Nexus for Exoplanet System Science, Virtual Planetary Laboratory Team, Box 351580, University of Washington, Seattle, Washington 98195, USA}

\author[0000-0001-7441-241X]{Celeste Hagee}
\affil{Department of Astronomy, University of Washington, Seattle, WA, 98195, USA}

\author[0000-0001-8187-0312]{Christiaan van Buchem}
\affil{University of Leiden}
\affil{Vrije Universiteit Amsterdam}

\author[0000-0002-2110-6694]{Matej Malik}
\affil{Department of Astronomy, University of Maryland, College Park, MD 20742, USA}

\author[0000-0002-2454-768X ]{Arjun B. Savel}
\affil{Department of Astronomy, University of Maryland, College Park, MD 20742, USA}

\author[0000-0002-7704-0153]{Matteo Brogi}
\affil{Department of Physics, University of Warwick, Coventry CV4 7AL, UK}
\affil{INAF—Osservatorio Astrofisico di Torino, Via Osservatorio 20, I-10025, Pino Torinese, Italy}
\affil{Centre for Exoplanets and Habitability, University of Warwick, Gibbet Hill Road, 
Coventry CV4 7AL, UK}

\author[0000-0003-3963-9672]{Emily Rauscher}
\affil{Department of Astronomy and Astrophysics, University of Michigan, Ann Arbor, MI, 48109, USA}

\author[0000-0002-1386-1710]{Victoria Meadows}
\affiliation{Department of Astronomy and Astrobiology Program, University of Washington, Box 351580, Seattle, Washington 98195, USA}
\affiliation{NASA Nexus for Exoplanet System Science, Virtual Planetary Laboratory Team, Box 351580, University of Washington, Seattle, Washington 98195, USA}

\author[0000-0003-4241-7413]{Megan Mansfield}
\affil{Steward Observatory, University of Arizona, Tucson, AZ 85719, USA}
\affil{NHFP Sagan Fellow}

\author[0000-0002-1337-9051]{Eliza M.-R.\ Kempton}
\affil{Department of Astronomy, University of Maryland, College Park, MD 20742, USA}

\author[0000-0002-0875-8401]{Jean-Michel Desert}
\affil{Anton Pannekoek Institute of Astronomy, University of Amsterdam, P.O. Box 94249, 1090GE Amsterdam, Noord Holland, NL}

\author[0000-0003-3191-2486]{Joost P. Wardenier}
\affil{Department of Physics (Atmospheric, Oceanic and Planetary Physics), University of Oxford, Oxford, OX1 3PU, UK}

\author[0000-0002-1321-8856]{Lorenzo Pino}
\affil{INAF - Osservatorio Astrofisico di Arcetri, Firenze, Italy}

\author[0000-0002-2338-476X]{Michael Line}
\affil{Steward Observatory, University of Arizona, Tucson, AZ 85719, USA}

\author[0000-0001-9521-6258]{Vivien Parmentier}
\affil{Department of Physics (Atmospheric, Oceanic and Planetary Physics), University of Oxford, Oxford, OX1 3PU, UK}

\author[0000-0003-4526-3747]{Andreas Seifahrt}
\affil{Department of Astronomy \& Astrophysics, University of
Chicago, Chicago, IL USA}

\author[0000-0003-0534-6388]{David Kasper}
\affil{Department of Astronomy \& Astrophysics, University of
Chicago, Chicago, IL USA}

\author[0000-0003-2404-2427]{Madison Brady}
\affil{Department of Astronomy \& Astrophysics, University of
Chicago, Chicago, IL USA}

\author[0000-0003-4733-6532]{Jacob L. Bean}
\affil{Department of Astronomy \& Astrophysics, University of
Chicago, Chicago, IL USA}

\begin{abstract}
Close-in lava planets represent an extreme example of terrestrial worlds, but their high temperatures may allow us to probe a diversity of crustal compositions. The brightest and most well-studied of these objects is 55 Cancri e, a nearby super-Earth with a remarkably short 17-hour orbit. However, despite numerous studies, debate remains about the existence and composition of its atmosphere. We present upper limits on the atmospheric pressure of 55 Cnc e derived from high-resolution time-series spectra taken with Gemini-N/MAROON-X. Our results are consistent with current crustal evaporation models for this planet which predict a thin $\sim$ 100 mbar atmosphere. We conclude that, if a mineral atmosphere is present on 55 Cnc e, the atmospheric pressure is below 100 mbar. 

\end{abstract}

\section{Introduction}\label{sec:Introduction}
Lava planets---worlds whose equilibrium temperature is so high that a significant portion of their crust is molten---represent the extreme in the canon of rocky terrestrial planets. 
55 Cancri e \citep{McArthur2004} is a canonical lava planet. It is a super-Earth with a mass of 8.63 M$_{\oplus}$ and density of 6.66 g cm$^{-3}$, comparable to, but higher than Earth's 5.5 g cm$^{-3}$. It has a radius of 2.00 R$_{\oplus}$, and a period of just 0.736 days \citep{Bourrier2018, Crida2018}. Spitzer phase curves suggest that 55 Cnc e is synchronously rotating \citep{Hammond2018}, with two different dayside brightness temperatures reported: \citet{Demory2016} reports that this value is 2700$_{-270}^{+270}$ K at 4.5 microns; a recent re-analysis performed by \citet{Mercier2022} finds it to be 3771$^{+665}_{-590}$ K. \citet{Demory2016} found a relatively large hotspot offset (which can indicate the presence of an advected atmosphere) of 
41$^{+12}_{-12}$ degrees in longitude, while the re-analysis of \citet{Mercier2022} indicates a smaller hotspot offset of -12$^{+21}_{-18}$ degrees.

In the last several years, many studies have used transmission spectroscopy to search for the presence and nature of 55 Cnc e’s atmosphere, with the growing consensus pointing towards a higher likelihood for either a high molecular weight atmosphere or no atmosphere.  Initial support for 55 Cnc e having a low mean molecular weight atmosphere, came from \citet{Tsiaras2016} who used HST/WFC3 data to infer a hydrogen-rich atmosphere and a tentative detection of HCN. \citet{Esteves2017} also found transmission signatures consistent with either an H$_2$-dominated, or high mean molecular weight atmosphere with H$_2$O, or terminator clouds.  However, subsequent observations failed to support the presence of an H$_2$-dominated atmosphere, with \citet{Deibert2021} using non-detections to place low constraints on HCN, NH$_3$, and C$_2$H$_2$, and \citet{Zhang2021} failing to detect the expected escaping He. \citet{Bourrier2018} also argued against the presence of an H/He envelope based on the bulk density and radius of the planet, and concluded that a high molecular weight atmosphere was likely present, in agreement with the results of \citet{Demory2016} and \citet{Angelo2017}. \citet{Jindal2020} subsequently used Gemini-N/GRACES and a sensitive non detection of TiO and water, to rule out low molecular weight or clear-sky water-rich atmospheres, although the results were also consistent with cloudy or no atmosphere. \citet{Ridden-Harper2016} and \citet{Keles2022} both used high-resolution transmission spectroscopy to search for traces of a mineral atmosphere but found none.

Photometry has also been used to to understand 55 Cnc e's thermal emission and search for signs of an atmosphere. \citet{Winn2011} used the Microvariability and Oscillations of STars (MOST) \citep{Rucinski2003} satellite to discover an unexplained IR phase modulation (i.e. a change in the sinusoidal brightness function associated with the changing phase) too large to be caused by combined star-planet fluctuations in brightness. Later, \citet{Sulis2019} used MOST to confirm that the thermal emission of the planet was variable through time. The IR Spitzer results from \citet{Demory2016} suggested that 55 Cnc e has either an optically thick atmosphere, or surface magma flow. However, a second analysis of the Spitzer phase curves found a higher dayside temperature at 4.5 $\mu$m, which was interpreted as emission from a temperature inversion generated in an atmosphere of UV-absorbing SiO vapor \citep{Mercier2022}.

Signatures of crustal evaporation in particular is an intriguing avenue of exploration for 55 Cnc e because of the predicted detectability of evaporated species such as sodium, silicate oxide, potassium, oxygen, iron and magnesium \citep{Ito2015, Ito2022}. At significant enough rates of evaporation, minerals with strong features in the optical spectrum should appear at the 10$^{-4}$ planet-to-star contrast ratio, a level observable by a large ground-based telescope equipped with a high-resolution spectrograph. Additionally, SiO and SiO$_2$ features may be detectable with the MIRI instrument on JWST \citep{Zilinskas2022}. 

Here we present an informative non-detection of gaseous iron in 55 Cnc e with optical dayside emission spectroscopy. In Section~\ref{sec:DataModels}, we discuss the observation and modeling of the planet. In Section~\ref{sec:Methods}, we describe our cross-correlation methodology and signal smearing mitigation strategy. In Section~\ref{sec:Results}, we present our results.

\section{Data \& Models}\label{sec:DataModels}

\begin{figure*}
\includegraphics[scale=.72,trim={2.5cm 6cm 5cm 2.5cm}]{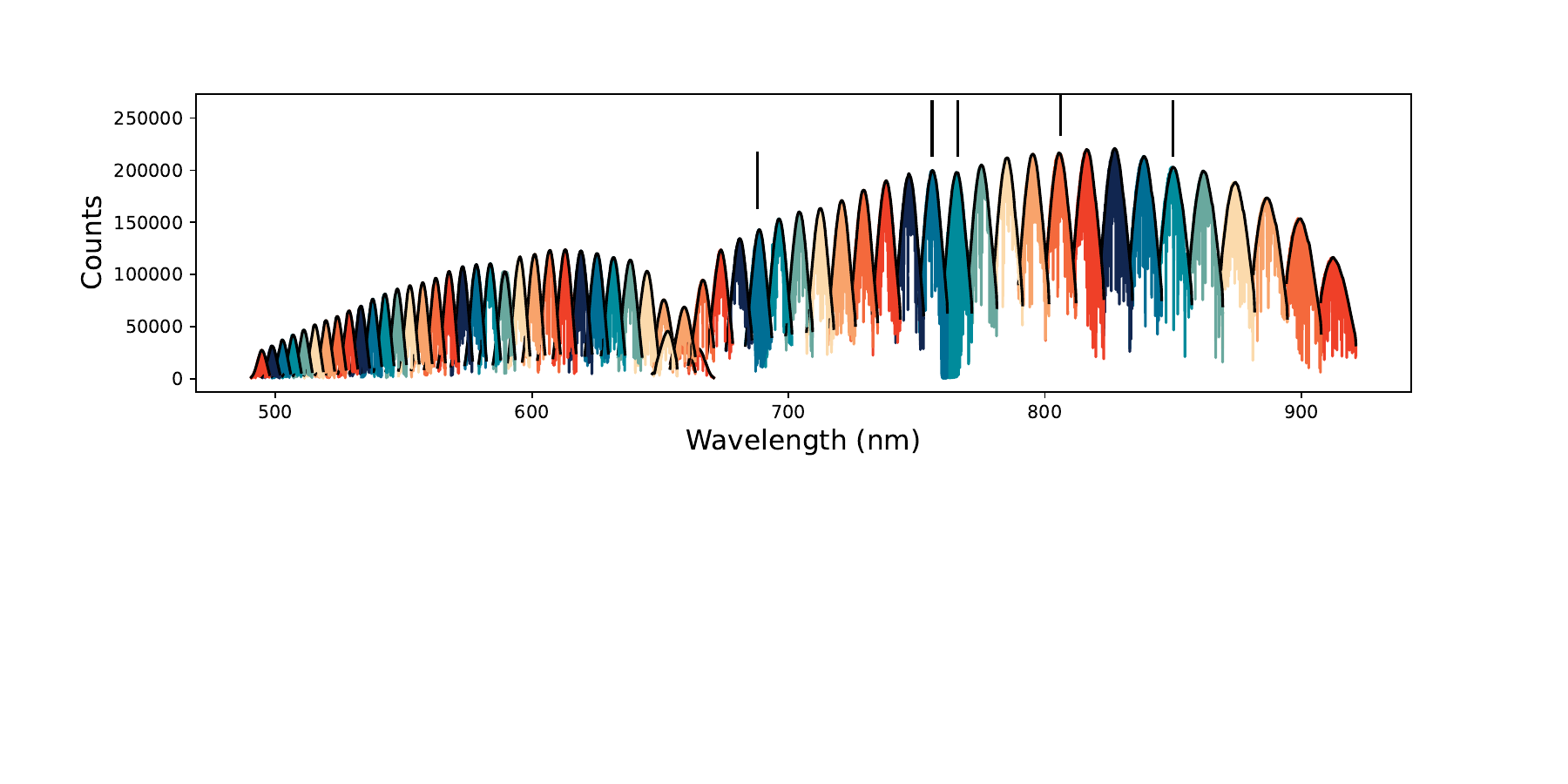}
\caption{One full set of echelle orders for one spectrum. The y-axis is raw counts. The x-axis is wavelength in nm. The blue SNR is $\sim$ 260 and the red SNR is $\sim$ 370. The orders indicated by the black lines were removed for having saturated lines or containing tellurics.} 
\label{fig:echelle}
\end{figure*}

Spectroscopic data were obtained at R $\sim$ 85,000 with the MAROON-X optical spectrograph on Gemini-North \citep{Seifahrt2016, Seifahrt2018, Seifahrt2020, Seifahrt2022}. The observation began at 07:48 UTC on March 27, 2022, and ended at 11:12 UTC. The planned exposure time for each frame was 130 s; however, the seeing on Maunakea was below average ($\sim$ 0.7 arcsec), prompting us to increase our exposure time to 300 s per frame for each of the 31 total frames we observed. The mean SNR of the blue frames was 260 and the mean SNR of the red frames was 370, as shown in Figure~\ref{fig:echelle}. We began the observation just after secondary eclipse ($\phi$ = 0.53) and ended it just before quadrature ($\phi$ = 0.73). We note that our seeing-driven increase in exposure time adversely affected the quality of the spectra; the exoplanet signal was subjected to ``smearing'', a poorly-documented phenomenon in which the exoplanet lines cross multiple resolution elements during an exposure, thus lowering the signal per resolution element. We discuss mitigation strategies for this ``smearing'' phenomenon in Section~\ref{sec:Methods}. 

\begin{figure}
\includegraphics[scale=.63,trim={1cm 0cm 0cm 0}]{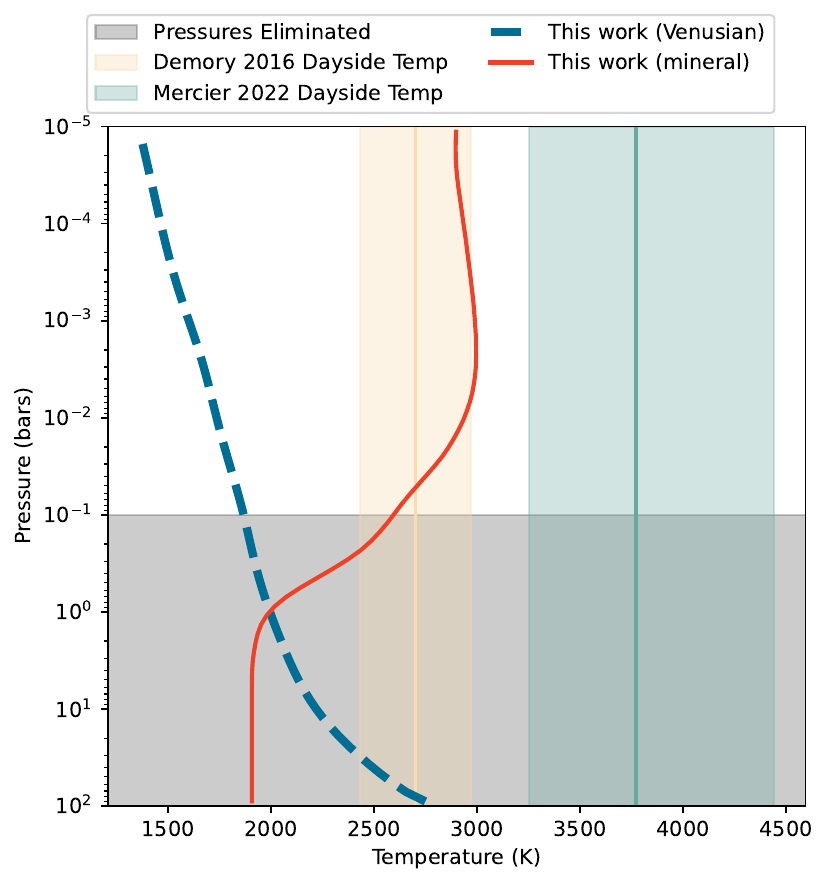}
\caption{Temperature-pressure profiles used in this work from the HELIOS code \citep{Malik2019b,Malik2019a}. The Venusian models were not detected in any configuration of our injection tests.} 
\label{fig:pt}
\end{figure}

\begin{figure}
\includegraphics[scale=.69, trim={1cm .5cm 0cm 0cm}]{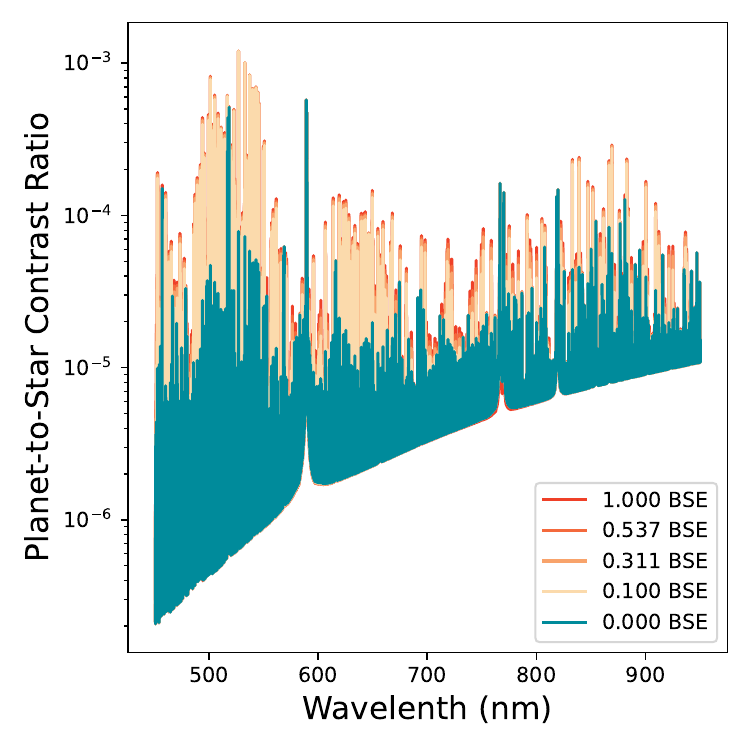}
\caption{Model spectra for all species inclusive. The planet-to-star contrast ratio for all 10 bar atmospheres at varying fractions of Fe of Bulk Silicate Earth (BSE) is shown; all other abundances are held constant.} 
\label{fig:models}
\end{figure}

We simulated the model emission spectra of 55 Cnc e with the radiative-convective code HELIOS \citep{Malik2019b,Malik2019a}. We assume dayside-averaged conditions in radiative-convective equilibrium with no day-to-night heat transport, motivated by the strong day-night contrast measured by \citet{Mercier2022} (T$_{day}$ = 3771$_{-520}^{+669}$ K, T$_{night}$ $<$ 1649 K at 2 $\sigma$) and by the short radiative timescales (vs advective transport timescales) expected for 55 Cnc e.  The atmospheric surface pressure is set to 10 bar, in between previous estimates of $\sim$ 1.4 - 200 bar from the photometric/spectroscopic studies of \citet{Angelo2017,Bourrier2018}, respectively.

We compute two model atmospheric grids using HELIOS. For the first grid, we calculated temperature--pressure profiles under silicate earth (BSE) composition \citep{Ito2015, Palme2003} and under a Venus-like, CO$_2$-dominated atmosphere. The BSE composition with varying fractions of Fe (atmospheric mixing ratios of 0.1, 0.311, 0.537, and 1.0) is motivated by the potential for an evaporated mineral atmosphere in magma ocean planets \citep{Schaefer2009, Miguel2011, Kite2016}. Because the composition includes a high abundance of silicate species with strong optical absorption, this model produces a strong temperature inversion. In contrast, the Venusian atmosphere is motivated by expectations for a solid planet \citep{Gaillard2014}, and its abundances produce a non-inverted temperature--pressure structure. These temperature--pressure profiles were then post-processed with varying Fe abundance, producing both pure-Fe and combined-species spectra based on the chemically consistent temperature--pressure profiles. 

Our second model grid solely is based on initial BSE abundances. However, we here take the additional step of modeling atmosphere--magma ocean interaction as detailed in \citet{vanBuchem2022} using an approach similar to that of MAGMA \citep{Fegley1987, Schaefer2004} and of \citet{Ito2015}. We assume chemical equilibrium between the magma and the overlying atmosphere, and the thermodynamics of the interior are modeled using the MELTS code \citep{Ghiorso1995}. Assuming evenly-mixed chemical profiles at the abundances produced by the atmosphere--magma ocean interaction at the dayside temperature of 55 Cnc e, we use HELIOS to post-process temperature--pressure profiles computed assuming BSE abundances. Because we do not compute vertical chemical abundances or link magma ocean output to atmospheric temperature--pressure profiles, we note that this method only provides an order-of-magnitude estimate for visibility in the high-resolution spectra. Additionally, we effectively vary the evenly mixed Fe abundance via the $\alpha_{\rm Fe}$ parameter, as described in \cite{putirka2019composition}. All temperature--pressure profiles can be seen in Figure~\ref{fig:pt} and spectra for the combination of all species in a 10 bar atmosphere can be seen in Figure~\ref{fig:models}.

Convolving our model spectra with the Spitzer Channel 2 filter response function, we find an eclipse depth of 91 ppm for our 100\% BSE model and 38 ppm for our Venusian model. Repeating this exercise in the CHEOPS band yields occultation depths of 5 ppm and 10 ppm for our BSE and Venusian models, respectively. While these models at first glance are discrepant from existing single Spitzer \citep{Mercier2022} and CHEOPS \citep{demory2022cheops} observed occultation depths of ${209}_{-47}^{+50}$ ppm and  $12 \pm 3$ ppm, respectively, the models fall within the full range of observed, variable occultations \citep{demory2016variability, demory2022cheops}. In light of the lack of consensus on the observed variability's origin, it is not immediately clear whether our models are constrained by existing photometric data. These values are 2.5$\sigma$ and 3.6$\sigma$ discrepant from the \cite{Mercier2022} analysis, respectively.
 
\section{Methods}\label{sec:Methods}

We use two established methods of exoplanet spectrum extraction to remove stellar and telluric lines. The principle behind both methods is the same: the exoplanet's lines rapidly shift over resolution elements, while the star's lines remain static to within one resolution element. Stationary features can thus be identified and removed via either Principal Component Analysis (PCA; \citet{Thatte2010}) or airmass detrending \citep{Brogi2019}. 

Because telluric contamination is not significant in the optical, we simply remove the handful of orders in which H$_{2}$O and O$_{2}$ are present. We then run PCA or airmass detrending. For PCA, which can operate in either the wavelength \citep{Zellem2014} or time regime \citep{DeKok2013}, common modes are identified between the input vectors and removed. The user can choose how many principal components are sufficient to maximize the detection of the exoplanet. For airmass detrending, the airmass trends of the spectrum are linearly fit and a median spectrum is divided out. This process is then repeated in the time domain. 

The products of the telluric removal process are then cross-correlated against an unaltered R $\sim$ 100,000 model spectrum, and the significance of that cross-correlation was assessed with the t-test method \citep{Brogi2019} to generate detection significances. In the case of PCA, we test every iteration from one to seven principal components. For airmass detrending we test the standard method as well as the modified methods described in \citet{Rasmussen2022}. To attempt to extract a signal from the spectra we use both methods with the understanding that to be robust, a significant cross-correlation peak at the expected planetary velocity should appear regardless of the method used. 

We also perform injection tests to determine which species we should detect. We tested Mg, SiO, Na, K for the mineral models and H$_{2}$O and CO$_{2}$ for the Venusian models. We note that the orders with O$_2$ were not considered in this analysis due to significant telluric contamination, so an O$_2$ injection test was not performed. This is accomplished by injecting the model spectra into the real data and putting the result through the cross-correlation code. An important caveat of this process is that the observed exoplanet signal was subjected to significant smearing during the observation due to the need for longer exposure times due to suboptimal seeing conditions (see Section~\ref{sec:DataModels}). The 300 s exposure times used in our observations were a factor of $\sim$ 10 greater than the typical time it takes for an exoplanet spectral line to cross one resolution element, thus we expect the signal to be smeared across $\sim 10$ resolution elements. 

We account for this smearing by ``top-hatting'' the model before injection tests. In this method, lines are first identified by fitting the continuum of the emission spectra with the spline interpolation algorithm SPORK (SPectral cOntinuum Refinement for telluriKs; \citet{Rasmussen2022}) and noting where features rise above it. The number of resolution elements crossed in the first exposure is then calculated from the change in radial velocity of the planet over time and the mean resolution of the spectrograph over its wavelength range using the equation: 

\begin{equation}
    \mathrm{N}_{\mathrm{RE}} = R\frac{\Delta v}{c},
\end{equation}
where $\mathrm{N}_{\mathrm{RE}}$ is the number of resolution elements crossed, $\Delta v$ is the change in radial velocity, R is the resolving power of the instrument, and c is the speed of light. 

\begin{figure}
\includegraphics[scale=.52,trim={2cm 3.5 4cm 1cm}]{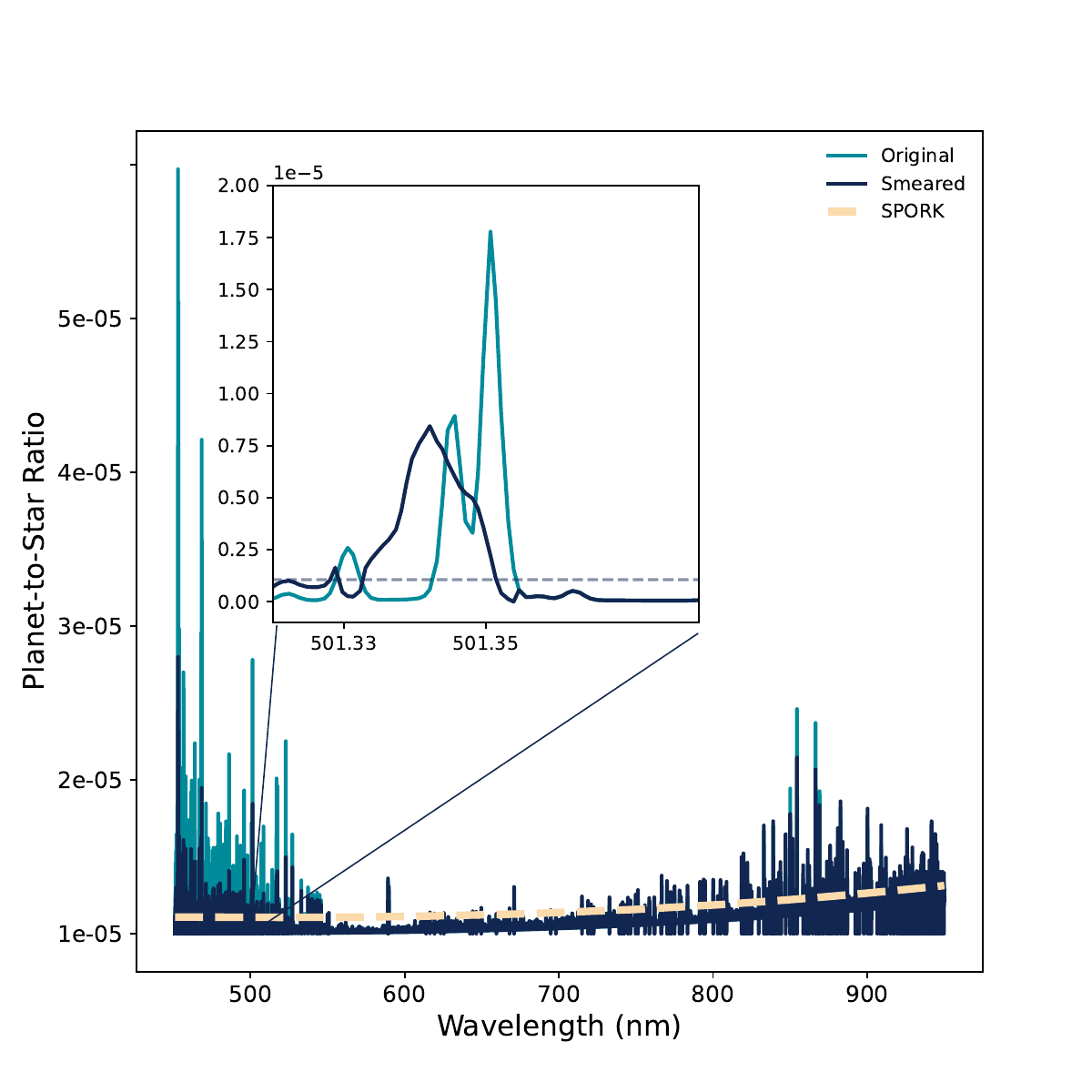}
\caption{``Top-hatting'' of the model. The continuum is fit with the spline interpolating function SPORK (cream colored in the main panel, grey in the inset window) described in \citet{Rasmussen2022}. The inset window shows an overlapping pair of emission lines which have been smeared out by the overlarge exposure time.} 
\label{fig:tophat}
\end{figure}

Each identified feature is then smeared by dividing its height (H) by the number of resolution elements crossed (N$_\mathrm{RE}$), and by adding that value to the next N$_\mathrm{RE}$ resolution elements. In this way the area under the emission line is conserved. A visualization of this process for a line-dense area of the spectrum is presented in Figure~\ref{fig:tophat}. We note a subtle difference in our ``top-hatting'' method compared to traditional boxcar smoothing: instead of smoothing equally about the center of a feature, our method smooths the signal toward shorter wavelengths, approximating smearing on the detector as the planet approaches quadrature. The top-hatted model is then used in our injection tests. Although the degree of smearing realistically differs from exposure to exposure, we use a constant smearing degree of 10 resolution elements (the maximum value), thus our estimate of our ability to recover the smeared signal is a conservative one. Finally, the injected data, after telluric removal, is cross-correlated against the un-smeared model. 

\section{Results and Discussion}\label{sec:Results}

We do not make any detections of any species in our data. We are, however, able to successfully rule out several scenarios. Iron is the only element which can be detected in our injection tests, so we focus on this element for our discussion.  Comparison of our injection tests with the data, are able to rule out heavier (P $>$ 100 mbar) BSE atmospheres to 4-$\sigma$, but not lighter ones. We are also unable to recover any signal above 4-$\sigma$ from the Venusian injection tests. However, it is possible that, without the signal smearing the data experienced, we would be sensitive to lower pressures. 

\begin{figure*}
\includegraphics[scale=.65,trim={5cm 0 5cm 0}]{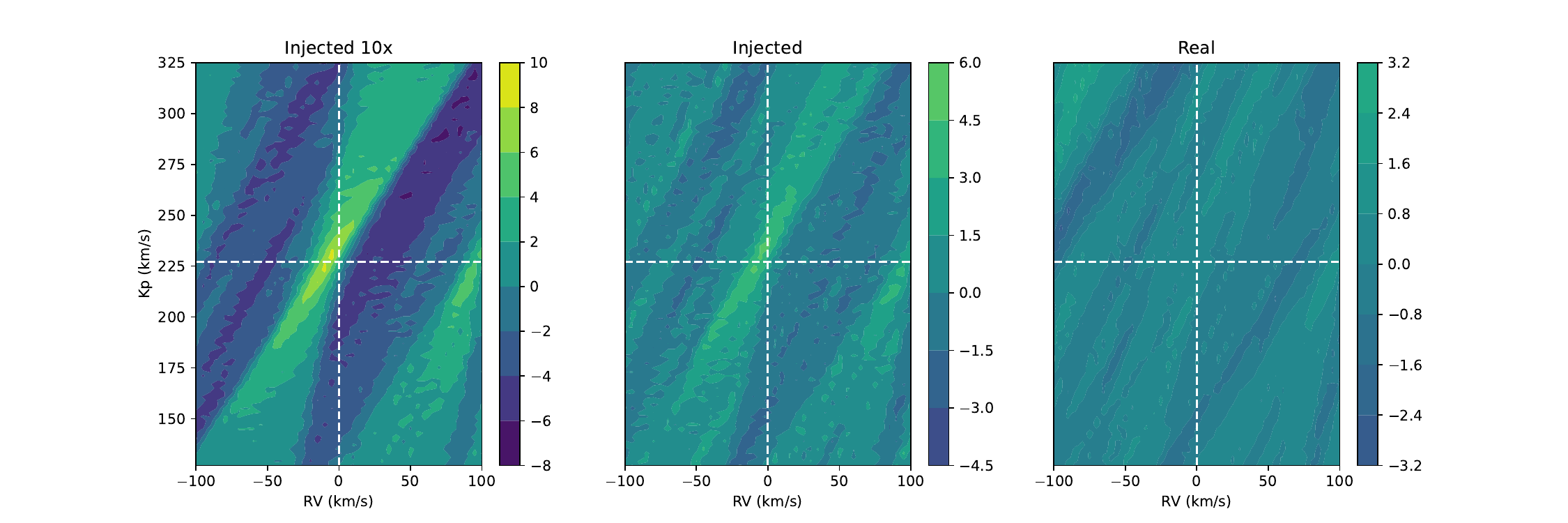}
\caption{The non-inverted 10 bar, 1.0 Bulk Silicate Earth (BSE) Fe atmosphere injection tests. (Left) The model is injected at 10 times the real contrast to illustrate where to expect the peak of the signal. The offset is due to the top-hatting (see Section \ref{sec:Methods}) of the model which does not center the smeared lines but offsets them to the blue. (Center) The injected signal at the correct contrast is recovered at $\sim$ 6.0 sigma. (Right) This model is not detected in the real dataset. } 
\label{fig:injections}
\end{figure*}

We perform injection tests for both our inverted and non-inverted modeled atmospheres, and several different Fe abundances (atmospheric mixing ratios of 0.1, 0.311, 0.537, and 1.0), and we are able to rule out all 100 mbar and 10 bar atmospheres at the 4-6 sigma level, as these should have been detectable in the data. Atmospheric pressure has the largest impact on detection, with Fe abundance producing a secondary effect---for example, peak detection significance decreases from 5.29 sigma to 3.82 sigma for 10 bar 1 BSE to 10 bar 0.1 BSE 
The 100 mbar atmospheres produces a 4.0 sigma detection, which we regard as ``tentative'' and the boundary for detection. At atmospheric pressures below this,  lower detection significances were obtained, which we deemed undetectable. An example of one of our injection tests is shown in Figure~\ref{fig:injections}: we first increase the contrast of the model atmosphere by a factor of ten in order to see where to expect the peak of the signal (left). Then we inject the planet at the real contrast ratio (center), and compare it to the injection-free data (right).

\citet{Mercier2022} claimed that a layer of emitting SiO is responsible for the high dayside temperature (3771\,K) detected with the re-analysis of Spitzer phase curves. Despite the relatively high SNR of our data, we are unable to test this hypothesis due to the relatively weak features of that species at such a high temperature. Future observations compiled with this one may be able to confirm or refute this.

Two different Spitzer analyses have yielded two different brightness temperatures ($\sim$ 2700 K from \citet{Demory2016} and $\sim$ 3700 K from \citet{Mercier2022}) for this planet. \citet{Demory2016}'s results are in line with our non-detections of Fe above P = 100 mbar (Figure~\ref{fig:pt}). The latter proposes that strong emission from an SiO band may drive the high brightness temperature derived in the Spitzer 4.5$\mu$m data. Both analyses yield deeper eclipse depths for 55 Cancri e than both our BSE and our Venusian atmospheric models. If 55 Cancri e has an atmosphere, this may imply that it has stronger temperature inversion than our models can produce with this set of chemical abundances.

\section{Conclusions}

We have obtained the first high-resolution, high-SNR emission spectrum of 55 Cnc e to investigate the presence of a mineral atmosphere on this lava planet. Although we do not detect any species, injection tests reveal that an atmosphere which includes any abundance of Fe should be detected down to 100 mbar. We thus conclude that if this hot, close-in planet possesses any atmosphere at all, it is likely the product of weak crustal evaporation and below the detectability threshold of our ground-based data. 55 Cnc e is a Cycle 1 JWST target; space-based observations at low- or medium-resolution may be able to confirm our results.

\acknowledgements
We thank our anonymous reviewer for their comments and suggestions that improved the clarity and robustness of the paper. K.C.R. and M.H.C. share co-first authorship. This work was performed by the Virtual Planetary Laboratory Team, a member of the NASA Nexus for Exoplanet System Science, funded via NASA Astrobiology Program Grant No. 80NSSC18K0829. This research was also supported by grant No. 2019-1403 from the Heising–Simons Foundation. 

\bibliography{bib}{}
\bibliographystyle{aasjournal}


\end{document}